\documentclass[aps,prc,twocolumn,showpacs,superscriptaddress]{revtex4}
\usepackage{graphicx,rotating,array,amsmath,amssymb,pifont}
\usepackage{dcolumn}

\begin{document}
\title{Half-life Limit of \nuc{19}{Mg}}
\author{N.~Frank}
  \affiliation{National Superconducting Cyclotron Laboratory,
               Michigan State University, East Lansing, MI, 48824}
  \affiliation{Department of Physics and Astronomy, Michigan State University,
               East Lansing, MI 48824}
\author{T.~Baumann}
  \affiliation{National Superconducting Cyclotron Laboratory,
               Michigan State University, East Lansing, MI, 48824}
\author{D.~Bazin}
  \affiliation{National Superconducting Cyclotron Laboratory,
               Michigan State University, East Lansing, MI, 48824}
\author{R.~R.~C.~Clement}
  \affiliation{National Superconducting Cyclotron Laboratory,
               Michigan State University, East Lansing, MI, 48824}
\author{M.~W.~Cooper}
  \affiliation{National Superconducting Cyclotron Laboratory,
               Michigan State University, East Lansing, MI, 48824}
\author{P.~Heckman}
  \affiliation{National Superconducting Cyclotron Laboratory,
               Michigan State University, East Lansing, MI, 48824}
  \affiliation{Department of Physics and Astronomy, Michigan State University,
               East Lansing, MI 48824}
\author{W.~A.~Peters}
  \affiliation{National Superconducting Cyclotron Laboratory,
               Michigan State University, East Lansing, MI, 48824}
  \affiliation{Department of Physics and Astronomy, Michigan State University,
               East Lansing, MI 48824}
\author{A.~Stolz}
  \affiliation{National Superconducting Cyclotron Laboratory,
               Michigan State University, East Lansing, MI, 48824}
\author{M.~Thoennessen}
  \affiliation{National Superconducting Cyclotron Laboratory,
               Michigan State University, East Lansing, MI, 48824}
  \affiliation{Department of Physics and Astronomy, Michigan State University,
               East Lansing, MI 48824}
\author{M.~S.~Wallace}
  \affiliation{National Superconducting Cyclotron Laboratory,
               Michigan State University, East Lansing, MI, 48824}
 \affiliation{Department of Physics and Astronomy, Michigan State University,
               East Lansing, MI 48824}
\email[]{frank@nscl.msu.edu}

\date{Draft: \today}

\begin{abstract}
A search for \nuc{19}{Mg} was performed using projectile fragmentation of
a 150~MeV/nucleon \nuc{36}{Ar} beam. No events of \nuc{19}{Mg} were
observed. From the time-of-flight through the fragment separator an
upper limit of 22~ns for the half-life of \nuc{19}{Mg} was established.
\end{abstract}


\maketitle

\section{Introduction}
\label{s:intro}

The search for the di-proton decay predicted over 40 years ago
\cite{Gol60} has recently intensified with the observation of
two-proton decay from \nuc{45}{Fe} \cite{Gio02,Pfu02}. These experiments
extracted the total energy of the decay, but they did not measure the
individual proton energies and angles. This information is necessary
in order to determine the nature of this decay, whether it is a
di-proton (\nuc{2}{He}) or a three-body decay.

Another promising candidate for di-proton decay is \nuc{19}{Mg}
\cite{Jan64,Gri00}. \nuc{19}{Mg} is predicted to be unbound by
0.9$\pm$0.3~MeV with respect to two-proton emission and bound by
0.66$\pm$0.5~MeV with respect to one-proton emission \cite{Aud93},
which is the ideal situation for the possibility of di-proton decay
\cite{Gol60}. The predictions of the two-proton decay energy are based
on extrapolations of the mass tables \cite{Aud93}. Microscopic
calculations have predicted that \nuc{19}{Mg} might even be bound depending on the choice of the force used in the calculation \cite{Des98}.  However, recent theoretical work by Grigorenko {\it et al.} suggests a half-life that is significantly smaller than the previously mentioned calculation \cite{Gri03}. 
Although a potentially very interesting nucleus, no
dedicated searches for the existence of \nuc{19}{Mg} have been performed
so far.

A search for the $T_z = -\frac{5}{2}$ nuclei \nuc{23}{Si}, \nuc{27}{S},
\nuc{31}{Ar}, and \nuc{35}{Ca}, discussed in Ref. \cite{Lan86}, should have
observed \nuc{19}{Mg} if it was indeed bound, however, it was not
observed and the implications of this are not mentioned. If
\nuc{19}{Mg} is unbound, the experimental requirements to study its
decay mode depend critically on the expected lifetime. The lifetime is
a strong function of the decay energy and the predicted range covers
many orders of magnitude.

We started a dedicated search for \nuc{19}{Mg} using projectile
fragmentation and extracted a first limit on its lifetime from the
time-of-flight through the fragment separator.

\section{Experimental Procedure}
\label{s:exp}

\begin{figure}
  \includegraphics[width=8cm]{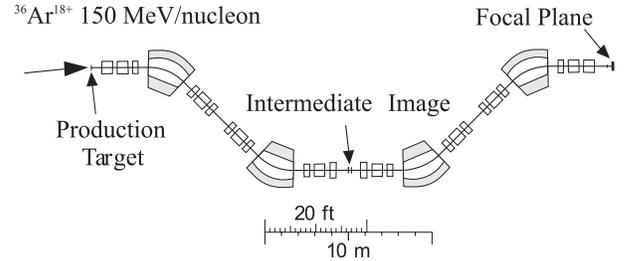}
  \caption{A1900 Fragment Separator Schematic.}
  \label{f:A1900}
\end{figure}

The experiment was performed at the Coupled Cyclotron Facility (CCF)
at the National Superconducting Cyclotron Laboratory (NSCL) at
Michigan State University. \nuc{36}{Ar} nuclei were accelerated by the
K500$\times$K1200 combination to 150~MeV/nucleon with a charge state of
18$^+$ and bombarded a 689~mg/cm$^2$ thick beryllium production
target. The average beam intensity was 4.3~pnA. The fragments of
interest were selected with the A1900 fragment separator \cite{Mor03}
shown in Figure \ref{f:A1900}. A 300-mg/cm$^2$-thick aluminum wedge
was placed at the intermediate image and the momentum acceptance was
limited to 0.5\%. The detectors at the focal plane of the separator
consisted of a 470-$\mu$m-thick $\Delta E$ silicon detector and a
10~cm thick plastic scintillator. The horizontal position acceptance
at the focal plane was limited to 15~mm in order to reduce the
background from lighter ions. The fragments were identified by their
energy loss in the silicon detector and the time-of-flight measured
between the plastic scintillator and a thin scintillator located at
the intermediate image. The total length of the A1900 from the
production target to the focal plane is 35.5~m.

\section{Data Analysis and Results}
\label{s:analysis}

\begin{figure}
  \includegraphics[width=8cm]{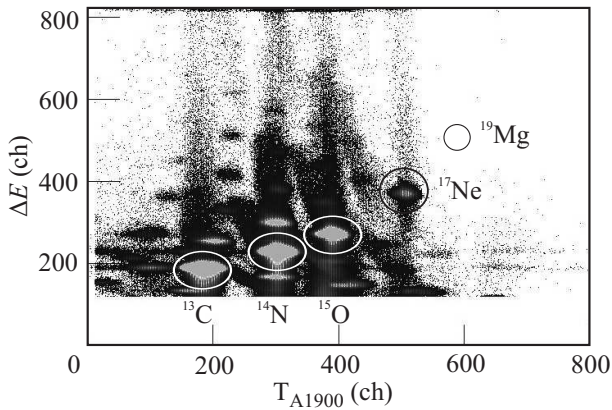}
  \caption{Energy loss versus time-of-flight identification spectrum for fragments reaching the focal plane.}
  \label{f:data}
\end{figure}

Figure \ref{f:data} shows the energy loss versus time-of-flight as
measured at the focal plane. The $N = 7$ isotones \nuc{13}{C}, \nuc{14}{N},
\nuc{15}{O}, and \nuc{17}{Ne} were clearly identified while no events for
\nuc{19}{Mg} were detected. Although contributions from pile-up,
incomplete charge collection, and other processes produced counts in
the spectrum, the region where \nuc{19}{Mg} is expected is background
free. This non-observation confirms that \nuc{19}{Mg} is indeed unbound
with respect to proton emission. It could be unbound with respect to
one- or two-proton emission, or both.

An upper limit of the half-life can be extracted from the predicted
number of produced \nuc{19}{Mg} and the transmission efficiency, and the
time of flight through the A1900 separator.

The fragment production of nuclei along the proton dripline has to be
extrapolated from the production of nuclei closer to the line of
stability and depends strongly on the reaction model. The most
commonly used approaches are the EPAX parameterizations
\cite{Sum90,Sum00} and calculations based on the abrasion-ablation
model \cite{Fri00}. In order to select the best reaction model for the
production of \nuc{19}{Mg}, we tuned the A1900 to maximize the production
of \nuc{20}{Mg}, which is particle stable. With the 689-mg/cm$^2$-thick
Be target, we measured a rate of 4~pps/pnA of \nuc{20}{Mg}.  The
simulation program LISE++ \cite{Baz02,Tar02} predicted rates of
5~pps/pnA, 10~pps/pnA, and 25~pps/pnA employing the EPAX1
\cite{Sum90}, EPAX2 \cite{Sum00}, and the abrasion-ablation model,
respectively. The fragment velocities and momentum distributions were
calculated following the parameterization of Morrissey
\cite{Mor89}. The EPAX1 calculation is within 20\% of the measured
\nuc{20}{Mg} rates and is the smallest prediction for the production of
very proton rich nuclei in this mass region for the given projectile
and target combination. Since the prediction by EPAX1 is the
smallest of the three approaches, they represent the most conservative
approach for extracting the upper limit of the half-life of \nuc{19}{Mg}.

It should be mentioned that $^{36}$Ar is not the optimum choice as a primary
beam. A more intense and pure beam of secondary beam of $^{20}$Mg can be 
produced from a primary beam of $^{24}$Mg \cite{Pie95}. However, this beam was not available for the present experiment.

EPAX1 yields a rate of 0.018~pps/pnA of \nuc{19}{Mg} to be transmitted to
the focal plane. We collected a total of 15.5~hours of beam on target
with an average beam intensity of 4.3~pnA. This would result in 4320
\nuc{19}{Mg} transmitted to the focal plane detectors if it was particle
stable.

\begin{figure}
  \includegraphics[width=8cm]{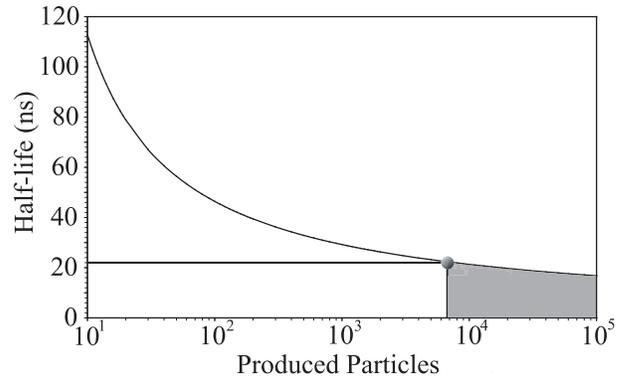}
  \caption{Lifetime as a function of \nuc{19}{Mg} transmitted to the focal plane.}
  \label{f:lifetime}
\end{figure}

The total flight path through the separator consists of two parts. In
the first half after the production target the average velocity of
\nuc{19}{Mg} is calculated to be $v = 0.46~c$. The velocity in the second
half is reduced to $v = 0.44~c$ following the 300~mg/cm$^2$ aluminum
wedge. The resulting time of flight is $T_\text{A1900} = 263$~ns.

The upper limit of the half-life assuming that one \nuc{19}{Mg} has not
decayed before reaching the focal plane was calculated using the
number of transmitted \nuc{19}{Mg} ($N$) and the total time-of-flight
($T_\text{A1900}$):

$$ T_{1/2} < T_\text{A1900} \frac{\ln 2}{\ln N}    .$$

This results in an upper limit of 22~ns for the half-life of
\nuc{19}{Mg}. The number of \nuc{19}{Mg} produced as predicted by LISE++ is
the largest uncertainty in the preceeding equation. However, Figure
\ref{f:lifetime} shows the relationship between the extracted limit on
the half-life and the predicted production. The half-life is not
strongly influenced by over- or under-predicting the rate of
\nuc{19}{Mg}. For instance, an over-prediction of the rate by a factor of
2 results in an upper half-life limit of 20~ns which is a difference of 10\%.

\begin{figure}
  \includegraphics[width=8cm]{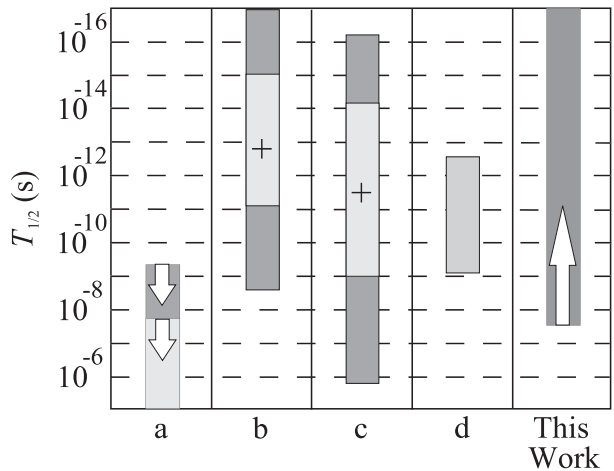}
  \caption{Theoretical calculations (a: Descouvemont \cite{Des98}), (b: Brown \cite{Bro02}), (c: Audi-Wapstra \cite{Aud93}), and (d: Grigorenko \cite{Gri03}) and experimental half-life ranges and limits.}
  \label{f:predictions}
\end{figure}

The upper half-life limit can be compared to recent three-body
calculations \cite{Gri00,Gri03}. In contrast to single proton
emitters, where the half-life can be directly related to the decay
energy through barrier penetration calculations \cite{Hof94}, the
situation for two proton emitters is more complicated.

Grigorenko {\it et al.} derived the half-life of \nuc{19}{Mg} as a
function of decay energy for three-body, di-proton and uncorrelated
two-proton decay \cite{Gri00}. Figure \ref{f:predictions} shows the
extracted half-life for the possible decay energies of Descouvement
($<$0.5~MeV) \cite{Des98}, Brown (1.1$\pm$0.3~MeV) \cite{Bro02}, and
Audi-Wapstra (0.9$\pm$0.3~MeV) \cite{Aud93}. The light shaded areas
correspond to the limits based on the decay energy uncertainties and
the three-body model of Ref. \cite{Gri00}, while the dark shaded areas
include the limits for the di-proton and uncorrelated two proton
decay. The fourth column of Figure \ref{f:predictions} shows the
half-life predictions from a refined three-body calculation by
Grigorenko {\it et al.} \cite{Gri03} and the last column is the
experimental limit extracted from the present work. The experimental
limit is consistent with the predictions. It rules out that \nuc{19}{Mg}
is particle-stable but does not put constraints on the calculations.

A different experimental approach is necessary to search for the
di-proton emission of \nuc{19}{Mg}. Simply increasing the production of
\nuc{19}{Mg} nuclei by extending the beam time or increasing the beam
intensity will not lead to a significant improvement as can be seen in
Figure \ref{f:lifetime}. Fragment separators are particularly useful
in this type of experiment because the fragmentation products are
cleanly separated and identified. In case no events are observed, it
is essential that the region where \nuc{19}{Mg} is expected to occur is
background free. However, the limit on the half-life is directly
proportional to the flight time. Therefore if one assumes that
\nuc{19}{Mg} exists before decay on the order of a few ns, an experiment
apparatus with a smaller flight time is necessary.

One possibility to reduce the time-of-flight that is currently pursued
is the use of single-neutron stripping reactions from a secondary beam
of \nuc{20}{Mg} to direcly observe the decay \cite{Muk01,Fra02}.

\section{Summary and Conclusions}
\label{s:conclusion}

The non-observation of \nuc{19}{Mg} from the fragmentation of \nuc{36}{Ar}
confirms the expected particle instability of this potential di-proton
emitter. An upper limit of the lifetime of 22~ns was established
for the first time. However, this limit is not yet sufficient to
constrain the experimental search for the di-proton decay of
\nuc{19}{Mg}. 

\section*{Acknowledgements}

This work has been supported by the National Science Foundation grant number PHY01-10253.


\begin{thebibliography}{99}
\bibitem{Gol60} V. I. Goldansky, Nucl. Phys. {\bf 19}, 482 (1960).
\bibitem{Gio02} J. Giovinazzo {\it et al.}, 
                Phys. Rev. Lett. {\bf 89}, 102501 (2002).
\bibitem{Pfu02} M. Pf\"utzner {\it et al.}, 
                Eur. Phys. J. {\bf A14}, 279 (2002).
\bibitem{Jan64} J. J\"anecke, Nucl. Phys. {\bf 61}, 326 (1964).
\bibitem{Gri00} L. V. Grigorenko, R. C. Johnson, I. G. Mukha, I. J. Thompson, 
                and M. V. Zhukov, Phys. Rev. Lett. {\bf 85}, 22 (2000).
\bibitem{Aud93} G. Audi and A. H. Wapstra, Nucl. Phys. {\bf A565}, 1 (1993).
\bibitem{Des98} P. Descouvemont, Phys. Lett. {\bf B437}, 7 (1998).
\bibitem{Gri03} L. V. Grigorenko, I. G. Mukha and M. V. Zhukov, 
                Nucl. Phys. {\bf A713}, 372 (2003).
\bibitem{Lan86} M. Langevin {\it et al.}, Nucl. Phys. {\bf A455}, 149 (1986).
\bibitem{Mor03} D. J. Morrissey, B. M. Sherrill, M. Steiner, A. Stolz and 
                I. Wiedenhoever, 
                Nucl. Instrum. and Methods, {\bf B204}, 90 (2003).
\bibitem{Sum90} K. S\"ummerer, W. Br\"uchle, D. J. Morrissey, M. Sch\"adel, 
                B. Szweryn, and Y. Weifan, Phys. Rev. C {\bf 42} 2546 (1990).
\bibitem{Sum00} K. S\"ummerer and B. Blank, Phys. Rev. C {\bf 61} 034607 (2000).
\bibitem{Fri00} W. A. Friedman, M. B. Tsang, D. Bazin, and W.G. Lynch, 
                Phys. Rev. C {\bf 62}, 064609 (2000). 
\bibitem{Baz02} D. Bazin, O. B. Tarasov, M. Lewitowicz, and O. Sorlin,
                Nucl. Instrum. and Methods, {\bf A482}, 314 (2002).
\bibitem{Tar02} O. B. Tarasov, D. Bazin, Preprint NSCL-MSU, MSUCL-1248, November 2002.
\bibitem{Mor89} D. J. Morrissey, Phys. Rev. C {\bf 39}, 460 (1989). 
\bibitem{Pie95} A. Piechaczek {\it et al.}, Nucl. Phys. {\bf A584}, 509 (1995).
\bibitem{Hof94} S. Hofmann, GSI Report GSI-93-04, 1993: {\it Handbook of Nuclear Decay Modes},
                (CRC Press, Boca Raton, FL, 1994).
\bibitem{Bro02} B. A. Brown, private communication.
\bibitem{Muk01} I. G. Mukha and G. Schrieder, 
                Nucl. Phys. {\bf A690}, 280c (2001).
\bibitem{Fra02} N. Frank {\it et al.}, Experiment proposal 02029, PAC26,
               NSCL MSU (2002). 
\end{thebibliography}
\end{document}